\documentclass{elsart}

\usepackage{amssymb}

\begin{document}

\begin{frontmatter}



\title{Comment on the article ``UCN anomalous losses and the UCN
capture cross section on material defects'' by A. Serebrov et al.}


\author[a]{A.L. Barabanov}
\ead{barab@mbslab.kiae.ru}
and
\author[b]{K.V. Protasov}
\ead{protasov@lpsc.in2p3.fr}

\address[a]{Kurchatov Institute, 1, Kurchatov square, Moscow 123182, Russia}
\address[b]{Laboratoire de Physique Subatomique et de Cosmologie, IN2P3-CNRS, UJFG, 53, Avenue des Martyrs, F-38026 Grenoble
Cedex, France}

\begin{abstract}
We present correct solution of the problem about a scattering of
the neutron on a point-like defect existing in a medium and show
that this mechanism cannot explain anomalous losses of UCN in
storage bottles.
\end{abstract}

\end{frontmatter}

In a recent article by Serebrov et al. \cite{Serebrov}, a possible
explanation of the problem of anomalous losses of UCN by
scattering on the defects was proposed. Unfortunately this
explanation is based on a wrong solution to the quantum mechanical
scattering problem and thus cannot be considered as a credible
hypothesis.

Serebrov et al. consider the problem on the scattering of neutrons
on a point-like defect existing in a medium. Following the article
\cite{Serebrov}, the medium bulk is described by usual Fermi
potential
\begin{equation}
\label{V1}
V_1(\mathbf{r})= \left\{ \begin{array}{ll}
U-iW, & z> 0, \\
0,& z < 0.
\end{array}
\right.
\end{equation}
The point-like defect situated at $\mathbf{r}_0$
($\mathbf{r}_0=(0,0,z_0)$ with $z_0 \geq 0$; $z_0=0$ corresponding
to the particular case of a defect on the surface of the medium)
is described by a potential
\begin{equation}
\label{V2} V_2= \frac{2\pi \hbar^2 B}{m} \delta
(\mathbf{r}-\mathbf{r}_0).
\end{equation}
The incident neutron propagates along the axis $z$ with the energy $E_0$.

According Serebrov et al. (see formulae (20) in \cite{Serebrov})
the cross-section of neutron capture by the defect is strongly
enhanced, i.e.
\begin{eqnarray*}
 \left.\sigma_{{\rm capt}}\right|_{E_0<U}=
\frac{|U-E_0|}{W} \left.\sigma_{{\rm capt}}\right|_{E_0>U},
\end{eqnarray*}
for subbarrier neutrons (indeed, typically $|U-E_0|/W\sim 10^4$).
This result is obviously wrong because the capture cross-section
for subbarrier neutrons appears to be infinite in a trivial limit
of non-absorbing medium ($W=0$).

The scattering problem can be easily solved within the well-known distorted
wave Born approximation (DWBA). This approximation is used in the
situation where a potential can be presented as a sum of two parts
$V=V_1+V_2$. The Schr\"odinger equation for the first potential
$V_1$ is solved exactly whereas the influence of the second
potential $V_2$ is taken into account in first order of the
perturbation theory.

The general answer for the scattering amplitude
$F^{(2)}(\mathbf{k}',\mathbf{k})$ can be found in any textbook on
scattering theory, for instance in \cite{Taylor}, and has the form
\begin{eqnarray}
\label{Solution} F^{(2)}(\mathbf{k}',\mathbf{k}) =
-\frac{m}{2\pi\hbar^2}\langle
\Psi^{(1)-}_{\mathbf{k}'}|V_2|\Psi^{(1)+}_\mathbf{k}\rangle,
\end{eqnarray}
where $\Psi^{(1)\pm}_\mathbf{k} (\mathbf{r})$ are exact solutions
of the Schr\"odinger equation for the first potential (distorted
waves). The difference with the usual Born approximation is in the
fact that the plane waves are replaced by exact solutions for the
first potential.

Exactly the same approach can be applied to calculate the capture
cross-section. In fact, the capture of the neutron by the defect
implies the transition from an initial state (propagation of
neutron up to capture $|i\rangle$, and $|0\rangle$ for radiation
field) to a final state (captured neutron $|f\rangle$ followed by
emission of $\gamma$ into the state $|\gamma\rangle$). The capture
cross-section is then given by the ratio of the capture rate to
the density flux of the incident neutrons. Using the Fermi rule,
we obtain
\begin{equation}
\label{eq4}
\sigma_{{\rm capt}}= \frac{2\pi m}{\hbar^2 k}\sum_{f,\gamma}
|\langle f,\gamma|\hat V|i,0\rangle|^2,
\end{equation}
where summation goes over all final neutron and $\gamma$ states,
and $\hat V$ is the coupling between the neutron-defect system and
the radiation field.

By direct analogy with (\ref{Solution}), formula (\ref{eq4}) gives
the capture cross section which is proportional to the squared
modulus of the initial neutron wave function at the defect,
\begin{equation}
\label{eq5}
\sigma_{{\rm capt}}=
|\Psi^{(1)+}_\mathbf{k}(\mathbf{r}_0)|^2
\sigma_{{\rm capt}}^0,
\end{equation}
where $\sigma_{{\rm capt}}^0$ is the capture cross section on the
isolated (in vacuum) defect. Here, only the initial state is
distorted by the potential (\ref{V1}).

In the problem considered in \cite{Serebrov}, the neutron wave
function can be easily calculated for the potential (\ref{V1})
and, in a particular case $E< U_0$, it has the form:
\begin{eqnarray*}
\Psi^{(1)+}_\mathbf{k}(\mathbf{r}_0) =
\frac{2k}{k+i\mbox{\ae}}e^{-\mbox{\ae} z_0}
\end{eqnarray*}
with $k = \sqrt{2mE_0/\hbar^2}$ and
$\mbox{\ae}=\sqrt{2m(U-iW-E_0)/\hbar^2}=\mbox{\ae}'-i\mbox{\ae}''$,
the neutron momenta outside and inside the medium. Thus the
capture cross section is of the form
\begin{eqnarray*}
\left.\sigma_{{\rm
capt}}\right|_{E_0<U}=
\frac{4k^2}{(k+\mbox{\ae}'')^2+\mbox{\ae}'^2}e^{-2\mbox{\ae}'z_0}\,
\sigma_{{\rm capt}}^0\simeq \frac{4E}{U}e^{-2\mbox{\ae}'z_0}\,
\sigma_{{\rm capt}}^0.
\end{eqnarray*}
The factor of proportionality takes the maximal value 4 at $E_0=U$
and $z_0=0$. It results from coherent summation of contributions
from the incident and reflected neutron waves.

For $E_0<U$ and for realistic distances $z_0>0$ between the defect
and the medium surface, the factor
$|\Psi^{(1)+}_\mathbf{k}(\mathbf{r}_0)|^2$ determines suppression
but not enhancement and has very simple physical interpretation.
The exponential term is an attenuation of the neutron wave
function in the medium. The deeper the defect inside the medium,
the smaller (exponentially) this factor.

To conclude notice that the phenomena of enhancement or
suppression of a process probability due to interaction in a final
and/or in a initial state are very well known in physics. As
example, let us cite only the case of Coulomb interaction in
nuclear reactions. Coulomb interaction can suppress the
cross-sections of nuclear reactions at low energies due to the
repulsion between nuclei or enhance the cross-sections and change
their behavior (from usual $1/v$-law to $1/v^2$-law) if the
charges of strongly interacting particles are opposite (attractive
Coulomb interaction) \cite{Wigner}. These examples illustrate
perfectly the physics of this phenomenon: an attractive
``external'' (long range) interaction pushes the wave function
into the region of ``internal'' short range) interaction and thus
enhances the latter cross-sections; a repulsive ``external''
interaction pushes the wave function out and suppresses the cross
sections. Therefore no repulsive interaction (as in the case
discussed in \cite{Serebrov} with positive Fermi potential) can
produce an enhancement phenomenon and explain anomalous losses of
UCN.



\end{document}